\documentclass[aps,prd,reprint,showpacs,longbibliography,groupedaddres,
titlepage,nofootinbib]{revtex4-1} 

\usepackage{amsthm}
\usepackage{amssymb}   
\usepackage{mathtools} 
\usepackage{hyperref}
\hypersetup{colorlinks=true,linkcolor=blue,urlcolor=blue,citecolor=blue}
\usepackage{accents}
\usepackage{tensor}
\usepackage[cal=boondox]{mathalfa}
\usepackage{lipsum}
\usepackage{relsize}
\usepackage{color}
\usepackage{comment}
\usepackage{nameref}
\usepackage[T1]{fontenc}
\newcommand{\ovg}[1]{\stackrel{(\gamma)}{#1}}
\newcommand{\uac}[1]{\underaccent{\tilde}{#1}}

\begin{document}
	
	\title{Canonical analysis of Holst action without second-class constraints}

	\author{Merced Montesinos\href{https://orcid.org/0000-0002-4936-9170} {\includegraphics[scale=0.05]{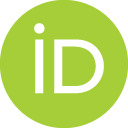}}}
	\email{merced@fis.cinvestav.mx}
	\author{Jorge Romero\href{https://orcid.org/0000-0001-8258-6647} {\includegraphics[scale=0.05]{ORCIDiD_icon128x128.png}}}%
	\email{ljromero@fis.cinvestav.mx}
		\author{Mariano Celada\href{https://orcid.org/0000-0002-3519-4736} {\includegraphics[scale=0.05]{ORCIDiD_icon128x128.png}}}
	\email[]{mcelada@fis.cinvestav.mx}
	\affiliation{%
		Departamento de F\'{i}sica, Cinvestav, Avenida Instituto Polit\'{e}cnico Nacional 2508, \\
		San Pedro Zacatenco, 07360 Gustavo A. Madero, Ciudad de M\'exico, Mexico	
	}%

	\date{\today}
	
	\begin{abstract}
		We perform the canonical analysis of the Holst action for general relativity with a cosmological constant without introducing second-class constraints. Our approach consists in identifying the dynamical and nondynamical parts of the involved variables from the very outset. After integrating out the nondynamical variables associated with the connection, we obtain the description of phase space in terms of manifestly $SO(3,1)$ [or $SO(4)$, depending on the signature] covariant canonical variables and first-class constraints only. We impose the time gauge on them and show that the Ashtekar-Barbero formulation of general relativity emerges. Later, we discuss a family of canonical transformations that allows us to construct new $SO(3,1)$ [or $SO(4)$] covariant canonical variables for the phase space of the theory and compare them with the ones already reported in the literature, pointing out the presence of a set of canonical variables not considered before. Finally, we resort to the time gauge again and find that the theory, when written in terms of the new canonical variables, either collapses to the $SO(3)$ ADM formalism or to the Ashtekar-Barbero formalism with a rescaled Immirzi parameter.
	\end{abstract}
	
	\maketitle
	
	\section{Introduction}\label{intro}
	
	In the first-order formalism, (real) general relativity is described by the Holst action~\cite{Holst}, which is made of the Palatini action coupled to the Holst term via the Immirzi parameter~\cite{Immirzi9700}. In vacuum (with or without a cosmological constant), this action reproduces exactly the same dynamics contained in the metric formulation of Einstein's theory obtained from the Einstein-Hilbert action as long as the orthonormal frame be nondegenerate. Outstandingly, the Holst action establishes the Lagrangian setting of the Ashtekar-Barbero variables~\cite{Barbero9505}, which constitute the building blocks of loop quantum gravity~\cite{Ashtekar0407,RovBook,Rovelli1106,ThieBook,perez2013}. Nevertheless, the derivation of these variables makes use of the so-called time gauge, which breaks the Lorentz group $SO(3,1)$ down to the $SO(3)$ subgroup. This gauge fixing avoids the introduction of second-class constraints, simplifying the resulting canonical theory at the expense of local Lorentz invariance. 
	
	Because Lorentz invariance plays a fundamental role in modern physics, there have been different approaches tackling the Lorentz-covariant canonical analysis of the Holst action. Nonetheless, those perspectives introduce second-class constraints, which are dealt with at the end either by using the Dirac bracket~\cite{Alexcqg1720} or by solving them explicitly~\cite{Barros0100,Montesinos1801,MontRomEscCel,MonRomCel_Revit}. Remarkably, in Refs.~\cite{Montesinos1801,MontRomEscCel} the second-class constraints were solved while preserving the manifest Lorentz invariance of the theory, obtaining different sets of canonical variables for the phase space of general relativity that is now described by first-class constraints only.
	
	In the standard approach, the second-class constraints are introduced due to a mismatch between the number of independent components of the Lorentz connection and those of the orthonormal field. It turns out that the spatial part of the connection corresponds to the configuration variables, and their canonically conjugate momenta are related to the spatial part of the orthonormal frame. Because the number of independent variables in the canonical momenta surpasses the number of components in the spatial part of the frame, one must add a quadratic constraint on the momenta in order to even things out. The Hamiltonian evolution of this constraint then generates a secondary constraint which, together with the former, makes up the set of second-class constraints of general relativity. Solving these constraints is what Refs.~\cite{Barros0100,Montesinos1801,MontRomEscCel,MonRomCel_Revit} are devoted to. 
	
	Alternatively, instead of real general relativity, one can move to the self-dual Palatini action~\cite{samuel,Jacobson198739,Jacobsoncqg5} (obtained from the Holst action by taking the Immirzi parameter equal to the imaginary unit), which involves the self-dual part of the Lorentz connection. This action allows us to derive a Hamiltonian formulation of the theory (the Ashtekar formalism~\cite{Ashtekar8709,Ashprl57.2244}) resorting neither to the introduction of second-class constraints nor to the time gauge~\cite{ashtekar1991lectures} (see also \cite{MontesinosGRG2001}), thus preserving Lorentz invariance; however, as implied by the use of self-dual variables, the formulation is complex and needs to be supplemented with reality conditions [the variables take advantage of the isomorphism between the Lie algebras of $SO(3,1)$ and  $SO(3,\mathbb{C})$]. 
	
	So, in order to avoid complex variables and work with real ones, the introduction of second-class constraints to preserve Lorentz invariance seems inexorable. Is that so? In this paper we show that it is possible to perform the canonical analysis of the Holst action without introducing neither second-class constraints nor gauge fixings spoiling Lorentz invariance. This is accomplished by providing a parametrization of the spatial part of the connection that separates its dynamical components from its nondynamical ones. The former equal in quantity to the number of components of the spatial part of the frame (which in turn are related to the canonical momenta), whereas the latter appear quadratically in the action and then can be integrated out via their own equation of motion. The resulting canonical theory, being manifestly Lorentz covariant, agrees with the one reported in Ref.~\cite{Montesinos1801}; outstandingly, the derivation is much simpler than the original one and the geometrical meaning of the variables is clearer.
	
	The outline of the paper is as follows. After establishing our conventions, we perform the 3+1 decomposition of the action in Sec.~\ref{analysis}, identifying the dynamical variables that make up the presymplectic structure. We then reparametrize the spatial part of the connection in terms of these variables and some additional fields that turn out to be auxiliary fields. We get rid of the latter and arrive at the canonical formulation of general relativity with manifest Lorentz invariance and first-class constraints only. Later, we discuss the time gauge in Sec.~\ref{TG} and the implementation of canonical transformations in Sec.~\ref{other}. To close the paper, we give some conclusions.
	
	{\it Conventions}. Spacetime indices are denoted by greek letters ($\mu,\nu,\dots$) so that points on the spacetime manifold $M$ are labeled by coordinates $\{x^{\mu}\}=\{t,x^{a}\}$, where $t$ is the time coordinate (we use the ``dot'' notation for time derivatives when possible) and latin letters at the beginning of the alphabet $(a,b,\ldots=1,2,3)$ denote spatial indices. We assume that $M$ has a topology $\mathbb{R}\times\Sigma$ and foliate it by constant time hypersurfaces $\Sigma_t$, each of which is diffeomorphic to some given orientable 3-manifold $\Sigma$ without boundary. The coordinates $\{x^a\}$ label points on $\Sigma_t$ and from now on we just write $\Sigma$ for any of these constant time hypersurfaces. Frame indices are associated with capital letters $I,J,\ldots=\{0,i\}$, for $i=1,2,3$. These indices are raised and lowered with the metric $(\eta_{IJ})=\text{diag}(\sigma,1,1,1)$, where $\sigma=-1\ (\sigma=+1)$ in the Lorentzian (Euclidean) case. The frame rotation group corresponds to the Lorentz group $SO(3,1)$ for $\sigma=-1$ or to the rotation group $SO(4)$ for $\sigma=+1$. The weight of a tensor is either indicated with the presence of a tilde over or below it, or mentioned somewhere else in the paper. The internal tensor $\epsilon_{IJKL}$ and the spacetime tensor density $\underaccent{\tilde}{\eta}_{\mu\nu\lambda\sigma}$ ($\tilde{\eta}^{\mu\nu\lambda\sigma}$) are totally antisymmetric and such that $\epsilon_{0123}=+1$ and $\underaccent{\tilde}{\eta}_{t123}=+1$ ($\tilde{\eta}^{t123}=+1$). In addition, we define the three-dimensional Levi-Civita symbols as $\underaccent{\tilde}{\eta}_{abc}:=\underaccent{\tilde}{\eta}_{tabc}$ ($\tilde{\eta}^{abc}:=\tilde{\eta}^{tabc}$) and $\epsilon_{ijk}:=\epsilon_{0ijk}$. The symmetrizer and antisymmetrizer are defined by $V_{(\alpha\beta)}:=(V_{\alpha\beta}+V_{\beta\alpha})/2$ and $V_{[\alpha\beta]}:=(V_{\alpha\beta}-V_{\beta\alpha})/2$, respectively. Furthermore, for an antisymmetric quantity $V_{IJ}$ we define its internal dual as $\ast V_{IJ}:=(1/2)\epsilon_{IJKL}V^{KL}$ and also the object $\stackrel{(\gamma)}{V} _{IJ}:=P_{IJKL}V^{KL}$ for
	\begin{equation}
		P_{IJKL}:=\eta_{I[K|}\eta_{J|L]}+\frac{1}{2\gamma}\epsilon_{IJKL},
	\end{equation}
	where $\gamma\neq0$ is the Immirzi parameter.\footnote{We assume $\gamma\neq\pm\sqrt{\sigma}$, which means that the self-dual and anti-self-dual sectors are excluded in our approach.} Its inverse is given by
	\begin{equation}
	(P^{-1})^{IJKL}=\frac{\gamma^2}{\gamma^2-\sigma}\left(\eta^{I[K|}\eta^{J|L]}-\frac{1}{2\gamma}\epsilon^{IJKL}\right)
	\end{equation}
	and satisfies $(P^{-1})^{IJKL}P_{KLMN}=\delta_{[M}^I\delta_{N]}^J$. ``$\wedge$'' and ``$d$'' stand for the wedge product of differential forms and the exterior derivative, correspondingly.
	
	\section{Canonical analysis}\label{analysis}
	
	In the first-order formalism, the orthonormal frame $e^I$ (assumed to be nondegenerate) and the $SO(3,1)$ [or $SO(4)$] connection $\omega^I{}_J$ are independent degrees of freedom that encode the gravitational field. In terms of them, the Holst action~\cite{Holst} for general relativity is given by
	\begin{eqnarray}\label{Holst}
		S[e,\omega] &=& \kappa\int_M\Biggl\{\left[\ast(e^I\wedge e^J)+\frac{\sigma}{\gamma}e^I\wedge e^J\right]\wedge F_{IJ}\notag\\
		&&-\frac{\Lambda}{12}\epsilon_{IJKL}e^I\wedge e^J\wedge e^K\wedge e^L\Biggr\},
	\end{eqnarray}
	where $F^I{}_J:=d\omega^I{}_J+\omega^I{}_K\wedge \omega^K{}_J$ is the curvature of $\omega^I{}_J$, which is compatible with the metric $\eta_{IJ}$, $d \eta_{IJ} - \omega^K{}_I \eta_{KJ} - \omega^K{}_J \eta_{IK}=0$ (and thus $\omega_{IJ}=-\omega_{JI}$); $\kappa$ is a constant related to Newton's constant and $\Lambda$ is the cosmological constant. Adapted to the spacetime foliation, the frame and the connection can be written as $e^I=e_t{}^I dt+e_a{}^I dx^a$ and $\omega^I{}_J=\omega_t{}^I{}_J dt+\omega_a{}^I{}_J dx^a$, respectively. 
	
	Let us introduce a vector $n^I$ with the following two properties at fixed $t$: $e_a{}^In_I=0$ and $n_In^I=\sigma$. Explicitly, this vector takes the form,
	\begin{equation}\label{vectn}
		n_I=\frac{1}{6\sqrt{q}}\epsilon_{IJKL}\tilde{\eta}^{abc}e_a{}^Je_b{}^Ke_c{}^L,
	\end{equation}
	where $q>0$ (of weight +2) is the determinant of the spatial metric $q_{ab}:=e_a{}^Ie_{bI}$, whose inverse metric is denoted by $q^{ab}$. The projector on the orthogonal plane to $n^I$ is given by
	\begin{equation}\label{projectorn}
		q^I{}_J:=q^{ab}e_a{}^Ie_{bJ}=\delta^I_J-\sigma n^In_J.
	\end{equation}
	Geometrically speaking, since at each point of $M$ the orthonormal frame $e^I$ can be thought of as an isomorphism between the coordinate basis and the orthonormal basis of tangent space, the vector $n^I$ corresponds to the normal vector to the hypersurface $\Sigma$ with respect to the latter basis; likewise, tangent vectors to $\Sigma$ are translated into orthogonal vectors to $n^I$. Thus, the splitting of the tangent space into the orthogonal and parallel parts to $n^I$ encodes the spacetime foliation.
	
	The 3+1 decomposition of the action \eqref{Holst} yields (we recall that all spatial boundary terms will be neglected because $\Sigma$ has no boundary)
	\begin{eqnarray}
		S&=& \kappa\int_{\mathbb{R}\times\Sigma}dtd^3x \biggl\{-2\tilde{\Pi}^{aI}n^J\partial_t\ovg{\omega}_{aIJ}+\omega_{tIJ}\tilde{\mathcal{G}}^{IJ}\notag\\
		&&+\frac{1}{\sqrt{q}}e_t{}^I\Bigl[2\tilde{\Pi}^a{}_I\tilde{\Pi}^{bJ}n^K\!\ovg{F}_{abJK}+n_I\Bigl(\tilde{\Pi}^{aJ}\tilde{\Pi}^{bK}\ovg{F}_{abJK}\notag\\
		&&-2\Lambda q\Bigr)\Bigr] \biggr\},\label{action1}
	\end{eqnarray}
	where $F_{ab}{}^I{}_{J}= \partial_a \omega_b{}^I{}_J - \partial_b \omega_a{}^I{}_J + \omega_a{}^I{}_K \omega_b{}^K{}_J - \omega_b{}^I{}_K \omega_a{}^K{}_J$ is the curvature of $\omega_a{}^I{}_J$, $dtd^3x$ is a shorthand for $dt\wedge dx^1\wedge dx^2\wedge dx^3$, and we have defined
	\begin{equation}\label{Pi}
		\tilde{\Pi}^{aI}:=\sqrt{q}q^{ab}e_b{}^I
	\end{equation}
	and
	\begin{equation}\label{Gauss}
		\hspace{-1mm}\tilde{\mathcal{G}}^{IJ}:=-2P^{IJ}{}_{KL}\left[\partial_a(\tilde{\Pi}^{aK}n^L)+2\omega_a{}^K{}_M\tilde{\Pi}^{a[M}n^{L]}\right].
	\end{equation}
	Notice that in~\eqref{action1}, both $\omega_{tIJ}$ and $e_t{}^I$ appear linearly in the action and thus play the role of Lagrange multipliers. It is customary to split the latter into the components parallel and orthogonal to $n^I$. Thus, we can express it as
	\begin{equation}\label{e_t}
	e_t{}^I=N n^I+N^ae_a{}^I,
	\end{equation}
	where $N$ is the lapse function and $N^a$ is the shift vector~\cite{Arnowitt2008}. The term of the action~\eqref{action1} involving $e_t{}^I$ then becomes the combination $-N^a\tilde{\mathcal{V}}_a-\uac{N}\tilde{\tilde{\mathcal{Z}}}$
	for
	\begin{subequations}
	\begin{eqnarray}
		\tilde{\mathcal{V}}_a &:=& -2\tilde{\Pi}^{bI}n^J\ovg{F}_{abIJ},\label{vector1}\\
		\tilde{\tilde{\mathcal{Z}}} &:=& -\sigma\tilde{\Pi}^{aI}\tilde{\Pi}^{bJ}\ovg{F}_{abIJ}+2\sigma\Lambda q,\label{scalar1}
	\end{eqnarray}
	\end{subequations}
	and $\uac{N}:=N/\sqrt{q}$.
	
	Let us introduce, for future purposes, the densitized metric $\underaccent{\tilde}{\underaccent{\tilde}{h}}_{ab}:=q^{-1}q_{ab}$, whose inverse $\tilde{\tilde{h}}^{ab}$ is given by $\tilde{\tilde{h}}^{ab}=\tilde{\Pi}^{aI}\tilde{\Pi}^b{}_I$; its determinant $h:=\det(\tilde{\tilde{h}}^{ab})$ has weight +4 and is related to $q$ by $h=q^2$. Relationship~\eqref{Pi} can be inverted to express $e_a{}^I$ in terms of $\tilde{\Pi}^{aI}$, yielding
	\begin{equation}
		e_a{}^I=h^{1/4}\uac{\uac{h}}_{ab}\tilde{\Pi}^{bI},\label{eaI}
	\end{equation}
	which allows us to express the vector $n_I$ as
	\begin{equation}\label{vectnpi}
	n_I=\frac{1}{6\sqrt{h}}\epsilon_{IJKL}\uac{\eta}_{abc}\tilde{\Pi}^{aJ}\tilde{\Pi}^{bK}\tilde{\Pi}^{cL}.
	\end{equation}
	In addition, we also define the connection $\nabla_a$ compatible with $e_a{}^I$ that satisfies
	\begin{equation}\label{covdev0}
		\nabla_a e_b{}^I:=\partial_ae_b{}^I-\Gamma^c{}_{ab}e_c{}^I+\Gamma_a{}^I{}_Je_b{}^J=0.
	\end{equation}
	These are 36 equations for 18 unknowns $\Gamma^c{}_{ab}\ (=\Gamma^c{}_{ba})$ and 18 unknowns $\Gamma_{aIJ}\ (=-\Gamma_{aJI})$. Their solution is the Christoffel symbol  $\Gamma^c{}_{ab}$ for the spatial metric $q_{ab}$ and 
	\begin{eqnarray}
		\Gamma_{aIJ} &=& q^{bc}e_{b[I|}\left(\partial_a e_{c|J]} -\partial_c e_{a|J]}\right) \notag \\
		&& +  \sigma q^{bc} e_{b[I}n_{J]}n_{K} \left( \partial_{a} e_c{}^K+\partial_{c} e_a{}^K\right) \notag \\
		&& +q^{bc}q^{df}e_{aK} e_{b[I|}e_{d|J]} \partial_f e_c{}^K,	\label{GaIJ0}
	\end{eqnarray}
	with $n_I$ given by~\eqref{vectn}.
	Notice that~\eqref{Pi} and~\eqref{covdev0} imply that $\nabla_a$ annihilates $\tilde{\Pi}^{aI}$ too,
	\begin{equation}
		\nabla_a \tilde{\Pi}^{bI}=\partial_a \tilde{\Pi}^{bI}+\Gamma^b{}_{ac}\tilde{\Pi}^{cI}-\Gamma^c{}_{ac}\tilde{\Pi}^{bI}+\Gamma_a{}^I{}_J\tilde{\Pi}^{bJ}=0.\label{covdev}
	\end{equation}
	In terms of $\tilde{\Pi}^{aI}$, the expression~\eqref{GaIJ0}  becomes
	\begin{eqnarray}
	\Gamma_{aIJ}  &=& \uac{\uac{h}}_{ab}\tilde{\Pi}^{c}{}_{[I|}\partial_{c} \tilde{\Pi}^{b}{}_{|J]} + \uac{\uac{h}}_{ab}\uac{\uac{h}}_{cd}\tilde{\Pi}^{c}{}_K \tilde{\Pi}^{b}{}_{[I}\tilde{\Pi}^{f}{}_{J]}\partial_{f} \tilde{\Pi}^{dK}  \notag \\ 
	& &+  \uac{\uac{h}}_{bc} \tilde{\Pi}^{b}{}_{[I|}\partial_{a} \tilde{\Pi}^c{}_{|J]} - \uac{\uac{h}}_{ab}\uac{\uac{h}}_{cd} \tilde{\Pi}^{b}{}_K \tilde{\Pi}^{c}{}_{[I}\tilde{\Pi}^{f}{}_{J]} \partial_{f} \tilde{\Pi}^{dK}  \notag \\
	& & - \sigma \uac{\uac{h}}_{ab} \tilde{\Pi}^{c}{}_{[I}n_{J]} n_{K} \partial_{c} \tilde{\Pi}^{bK}  + \sigma \uac{\uac{h}}_{bc}\tilde{\Pi}^{b}{}_{[I}n_{J]}n_{K} \partial_{a} \tilde{\Pi}^{cK}, \notag \\ \label{GaIJ}	
	\end{eqnarray}	
	with $n_I$ given by~\eqref{vectnpi}. The curvature of $\Gamma_a{}^I{}_J$ is $R_{ab}{}^I{}_J = \partial_{a}\Gamma_b{}^I{}_J-\partial_{b}\Gamma_a{}^I{}_J + \Gamma_a{}^I{}_K \Gamma_{b}{}^K{}_J - \Gamma_b{}^I{}_K \Gamma_{a}{}^K{}_J$.

	Note that the first term on the right-hand side of the action~\eqref{action1} involves the time derivative of $\stackrel{(\gamma)}{\omega}_{aIJ}$ and thus contributes to the presymplectic structure of the theory. Since there are 18 independent components in $\stackrel{(\gamma)}{\omega}_{aIJ}$ (the same as in $\omega_{aIJ}$), the usual approach requires us to introduce the same number of canonically conjugate momenta. However, since these momenta are built up from the 12 components $e_a{}^I$, six additional constraints on the momenta must be added~\cite{Barros0100,Montesinos1801,MontRomEscCel,MonRomCel_Revit}. This is the traditional path taken, and it leads to the emergence of second-class constraints; one set being the aforementioned constraints on the momenta and the other arising from the preservation under time evolution of the former. Here, we will follow a different path inspired by our previous work~\cite{Montesinos1801}; instead of introducing constraints on the momenta, we will properly parametrize the 18 components of $\omega_{aIJ}$ into 12 dynamical and six nondynamical variables. The 12 dynamical variables thus will correspond to the configuration variables, whereas the six nondynamical ones will be later integrated out from the action, obtaining at the end a manifestly $SO(3,1)$ [or $SO(4)$] covariant canonical formulation of the Holst action. To do this, note that the presymplectic structure in~\eqref{action1} can be expressed as\footnote{From now on, $e_a{}^I$ appears no more; it is replaced by~\eqref{eaI}, and thus, $n_I$ is given by~\eqref{vectnpi}.}
	\begin{equation}\label{symplectic}
		-2\tilde{\Pi}^{aI}n^J\partial_t\ovg{\omega}_{aIJ}=2\tilde{\Pi}^{aI}\dot{C}_{aI},
	\end{equation}
	where we have introduced the 12 configuration variables $C_{aI}$ defined by [see Eq. (8) of Ref.~\cite{Montesinos1801}]
	\begin{equation}\label{defC}
		C_{aI}:=W_a{}^b{}_{IJK}\ovg{\omega}_b{}^{JK},
	\end{equation}
	with $W_a{}^b{}_{IJK}$ ($=-W_a{}^b{}_{IKJ}$) given explicitly by
	\begin{equation}\label{W}
		W_a{}^b{}_{IJK}:=-\left(\delta_a^b\eta_{I[J}n_{K]}+n_I\uac{\uac{h}}_{ac}\tilde{\Pi}^c{}_{[J}\tilde{\Pi}^b{}_{K]}\right), 
	\end{equation}
	which can be thought of as an operator that singles out the 12 configuration variables $C_{aI}$--constructed out of the components of the connection $\omega_{aIJ}$--that contribute to the resulting canonical symplectic structure. The expression \eqref{defC} can be solved for $\ovg{\omega}_{aIJ}$ to express the connection in terms of $C_{aI}$ plus six additional variables $\uac{\lambda}_{ab}\ (=\uac{\lambda}_{ba})$ living in the kernel of $W_a{}^b{}_{IJK}$,
	\begin{equation}\label{connec_par}
		\ovg{\omega}_{aIJ}=M_a{}^b{}_{IJK}C_b{}^K+\uac{\lambda}_{ab}\tilde{N}^b{}_{IJ},
	\end{equation}
	with $M_a{}^b{}_{IJK}\ (=-M_a{}^b{}_{JIK})$ and $\tilde{N}^b{}_{IJ}\ (=-\tilde{N}^b{}_{JI})$ being given by
	\begin{subequations}
	\begin{eqnarray}
		M_a{}^b{}_{IJK} &:=& 2\sigma\delta_a^b n_{[I}\eta_{J]K}+\sigma\uac{\uac{h}}_{ac}\tilde{\Pi}^c{}_{[I}\tilde{\Pi}^b{}_{J]}n_K \notag\\
		&& -\frac{\sigma}{2\gamma}\delta_a^b\epsilon_{IJKL}n^L  \notag \\
		&& -\frac{\sigma}{2\gamma}\epsilon_{IJMN}\uac{\uac{h}}_{ac}\tilde{\Pi}^{bM}n^N\tilde{\Pi}^c{}_K,\label{M} \\
		\tilde{N}^a{}_{IJ} &:=& \epsilon_{IJKL}\tilde{\Pi}^{aK}n^L.\label{N}
	\end{eqnarray}
	\end{subequations}
	[Confront with Eqs. (6), (9), and (10) of Ref.~\cite{Montesinos1801}]. In addition, we introduce the tensor density $\uac{U}_{ab}{}^{cIJ}\ (=\uac{U}_{ba}{}^{cIJ}=-\uac{U}_{ab}{}^{cJI})$ defined as
	\begin{equation}\label{U}
		\uac{U}_{ab}{}^{cIJ}:=\left(1-\frac{\sigma}{\gamma^2}\right)\ast(P^{-1})^{IJKL}  \delta^c{}_{(a} \uac{\uac{h}}_{b)e}\tilde{\Pi}^{e}{}_K n_L,
	\end{equation}
	where the internal dual in~\eqref{U} acts on either the first pair or the last pair of indices of $P^{-1}$ ($\ast$ and $P^{-1}$ commute with one another). Together, the objects \eqref{W}, \eqref{M}, \eqref{N} and \eqref{U} satisfy the following orthogonality relations:
	\begin{subequations}
	\begin{eqnarray}
		 W_a{}^{cIMN}M_c{}^b{}_{MNJ} &=& \delta_a^b\delta_J^I,\\
		\uac{U}_{ab}{}^{cIJ}\tilde{N}^d{}_{IJ} &= & \delta^{(c}{}_a \delta^{d)}{}_b,\\
		W_a{}^{(b}{}_{IJK}\tilde{N}^{c)JK} &=& 0,\\
		\uac{U}_{ab}{}^{cIJ}M_c{}^d{}_{IJK}&=& 0,
	\end{eqnarray}
	\end{subequations}
	as well as the completeness relation,
	\begin{equation}
		M_a{}^c{}_{IJM}W_c{}^{bMKL}+\tilde{N}^c{}_{IJ}\uac{U}_{ac}{}^{bKL}=\delta_a^b\delta_{[I}^K\delta_{J]}^L.
	\end{equation}
	Therefore, $W$ and $\uac{U}$ are orthogonal projectors that allow us to split the 18 components of the connection $\omega_{aIJ}$ [or $\ovg{\omega}_{aIJ}$] into the 12+6 variables $(C_{aI},\uac{\lambda}_{ab})$.  The associated decomposition given in~\eqref{connec_par} is induced by the canonical symplectic structure given by the right-hand side of~\eqref{symplectic}. The inverse of the map \eqref{connec_par} is given by~\eqref{defC} together with
	\begin{equation}\label{deflambda}
		\uac{\lambda}_{ab}=\uac{U}_{ab}{}^{cIJ}\ovg{\omega}_{cIJ},
	\end{equation}
which clearly shows that $C_{aI}$ and $\uac{\lambda}_{ab}$ are independent variables among themselves.

	Substituting~\eqref{connec_par} into~\eqref{Gauss}, \eqref{vector1} and \eqref{scalar1}, we obtain, after some algebra,
	\begin{widetext}
		\begin{subequations}
		\begin{eqnarray}
		\tilde{\mathcal{G}}^{IJ} &=& 2\tilde{\Pi}^{a[I}C_a{}^{J]}+4P^{IJ}{}_{KL}\tilde{\Pi}^{a[K}n^{M]}\Gamma_a{}^L{}_M,\label{Gauss2}\\
		\tilde{\mathcal{V}}_a &=& 2 \left( 2\tilde{\Pi}^{bI}\partial_{[a}C_{b]I}-C_{aI}\partial_b\tilde{\Pi}^{bI} \right) + (P^{-1})_{IJKL} \tilde{\mathcal{G}}^{IJ} 
		\left (M_a{}^{bKLM}C_{bM}+ \uac{\lambda}_{ab}\tilde{N}^{bKL}\right ),\label{vector2}\\ 
		\tilde{\tilde{\mathcal{Z}}} &=& -\sigma\tilde{\Pi}^{aI}\tilde{\Pi}^{bJ}R_{abIJ} +2\tilde{\Pi}^{a[I|}\tilde{\Pi}^{b|J]} \left [ C_{aI}C_{bJ}+2C_{aI}\ovg{\Gamma}_{bJK}n^K+\left(\Gamma_{aIL}+\frac{2}{\gamma}\ast\Gamma_{aIL}\right)\Gamma_{bJK}n^Kn^L  \right. \notag\\
		&& \left. +\frac{1}{\gamma^2}q^{KL}\Gamma_{aIK}\Gamma_{bJL} \right ] + 2\sigma\Lambda \sqrt{h}+2\tilde{\Pi}^{aI} n^J\nabla_a\tilde{\mathcal{G}}_{IJ}-\frac{1}{4}\left[\tilde{\mathcal{G}}^{IJ}-(P^{-1})^{IJKL}\tilde{\mathcal{G}}_{KL}+ 2\sigma n^I \tilde{\mathcal{G}}^J{}_K n^K \right]\tilde{\mathcal{G}}_{IJ}\notag\\
		&& +\frac{\sigma\gamma^2}{\gamma^2-\sigma}G^{abcd}\bigl(\uac{\lambda}_{ab}-\uac{U}_{ab}{}^{eIJ}\ovg{\Gamma}_{eIJ}\bigr)\bigl(\uac{\lambda}_{cd}-\uac{U}_{cd}{}^{fKL}\ovg{\Gamma}_{fKL}\bigr),\label{scalar2}
		\end{eqnarray}
	\end{subequations}
	\end{widetext}
	where the coefficients $G^{abcd}:=\tilde{\tilde{h}}^{ab}\tilde{\tilde{h}}^{cd}-\tilde{\tilde{h}}^{(a|c}\tilde{\tilde{h}}^{|b)d}$ have weight +4. Notice that there are no terms involving $\uac{\lambda}_{ab}$ in~\eqref{Gauss2}, and that~\eqref{vector2} and~\eqref{scalar2} depend on $\uac{\lambda}_{ab}$ but not on their derivatives. This is staggering, since the original expressions~\eqref{vector1} and \eqref{scalar1} involve derivatives of the connection. The action~\eqref{action1} now takes the suggestive form,
	\begin{eqnarray}
	S=&&\kappa\int_{\mathbb{R}\times\Sigma}dtd^3x \left ( 2\tilde{\Pi}^{aI}\dot{C}_{aI}+\omega_{tIJ}\tilde{\mathcal{G}}^{IJ}-N^a\tilde{\mathcal{V}}_a-\uac{N}\tilde{\tilde{\mathcal{Z}}} \right ), \notag\\
	&& \label{action2}
	\end{eqnarray}
	which really resembles what is expected when casting an action in Hamiltonian form. Nevertheless, we have not finished yet because the action~\eqref{action2} still depends on the variables $\uac{\lambda}_{ab}$ as indicated above. Notice that the map from $\omega_a{}^I{}_J$ to $C_{aI}$ and $\uac{\lambda}_{ab}$ through~\eqref{defC} and~\eqref{deflambda}, with an inverse map given by~\eqref{connec_par}, can be seen as a change of variables. Nevertheless, as is clear from~\eqref{symplectic} and~\eqref{defC}, the presymplectic structure present in~\eqref{action1} becomes the canonical symplectic structure present in~\eqref{action2} when such a map is used. Therefore, we reach a smaller phase-space and simultaneously parametrize it with manifestly Lorentz-covariant canonical variables ($C_{aI}$, $\tilde{\Pi}^{aI}$). The reduction map is given by $(\omega_a{}^I{}_J,  \tilde{\Pi}^{aI}) \longmapsto (C_{aI}, \tilde{\Pi}^{aI})$ using~\eqref{defC}. This reduction process leaves the null directions of the presymplectic structure~\eqref{action1} out of the canonical symplectic structure present in~\eqref{action2}. The null directions are clearly along $\uac{\lambda}_{ab}$, which turn out to be auxiliary fields~\cite{HennBook} that can be integrated out from the action by using their own equation of motion. In fact, instead of considering $\uac{\lambda}_{ab}$ as configuration variables, which would lead us to define their canonically conjugate momenta $\tilde{\tilde p}^{ab}$ and also to introduce second-class constraints in an enlarged phase space, we will take advantage of the fact that the variables $\uac{\lambda}_{ab}$ are auxiliary fields that can be integrated out by setting the variational derivative of the action~\eqref{action2} with respect to $\uac{\lambda}_{ab}$ equal to zero (this amounts to explicitly solving the second-class constraints that otherwise would appear), namely,
	\begin{eqnarray}
	&& \frac{2\sigma\gamma^2}{\gamma^2-\sigma}\uac{N}G^{abcd}\bigl(\uac{\lambda}_{cd}-\uac{U}_{cd}{}^{fIJ}\ovg{\Gamma}_{fIJ}\bigr)\notag\\
	&&+ (P^{-1})_{IJKL} N^{(a}\tilde{N}^{b)IJ}\tilde{\mathcal{G}}^{KL}=0.\label{eqauxfields}
	\end{eqnarray}
	This equation is linear in $\uac{\lambda}_{ab}$ and can be solved for them as long as $\uac{N}\neq0$, which is always fulfilled since the orthonormal frame was assumed to be nondegenerate. Hence, the solution for $\uac{\lambda}_{ab}$ is
	\begin{eqnarray}
		\uac{\lambda}_{ab} &=& \uac{U}_{ab}{}^{cIJ}\ovg{\Gamma}_{cIJ}\notag\\
		&&-\frac{\sigma(\gamma^2-\sigma)}{2\gamma^2\uac{N}}(G^{-1})_{abcd}N^c(P^{-1})_{IJKL}\tilde{N}^{dIJ}\tilde{\mathcal{G}}^{KL}, \notag\\
		\label{fixlambd}
	\end{eqnarray}
	with $(G^{-1})_{abcd} = (1/2)(\uac{\uac{h}}_{ab}\uac{\uac{h}}_{cd}-2\uac{\uac{h}}_{(a|c}\uac{\uac{h}}_{|b)d})$ of weight -4 [and thus $G^{abef}(G^{-1})_{cdef}= \delta^a{}_{(c} \delta^b{}_{d)}$]. Substituting~\eqref{fixlambd} back into the action \eqref{action2}, integrating by parts the term involving the covariant derivative in~\eqref{scalar2}, and collecting all the terms proportional to $\tilde{\mathcal{G}}^{IJ}$, the action acquires the final form,
	\begin{eqnarray}
	S=&&\kappa\int_{\mathbb{R}\times\Sigma}dtd^3x \left ( 2\tilde{\Pi}^{aI}\dot{C}_{aI}-\lambda_{IJ}\tilde{\mathcal{G}}^{IJ}-2N^a\tilde{\mathcal{D}}_a-\uac{N}\tilde{\tilde{\mathcal{H}}} \right ), \notag\\
	&& \label{action3}
	\end{eqnarray}
	where $\tilde{\mathcal{G}}^{IJ}$ is the same as in~\eqref{Gauss2}, whereas 
	\begin{subequations}
		\begin{eqnarray}
		\tilde{\mathcal{D}}_a &:=& 2\tilde{\Pi}^{bI}\partial_{[a}C_{b]I}-C_{aI}\partial_b\tilde{\Pi}^{bI},\label{diff}\\
		\tilde{\tilde{\mathcal{H}}} &:=& -\sigma\tilde{\Pi}^{aI}\tilde{\Pi}^{bJ}R_{abIJ} +2\tilde{\Pi}^{a[I|}\tilde{\Pi}^{b|J]}\Biggl[C_{aI}C_{bJ}\nonumber\\
		&& + 2C_{aI}\ovg{\Gamma}_{bJK}n^K+\left(\Gamma_{aIK}+\frac{2}{\gamma}\ast\Gamma_{aIK}\right)\Gamma_{bJL}n^Kn^L\nonumber\\
		&& +\frac{1}{\gamma^2}q^{KL}\Gamma_{aIK}\Gamma_{bJL}\Biggr]+2\sigma\Lambda \sqrt{h},\label{scalar3}
		\end{eqnarray}
	\end{subequations}
	and
	\begin{eqnarray}
                \lambda_{IJ} &:=& -\omega_{tIJ}-2\tilde{\Pi}^a{}_{[I}n_{J]}\nabla_a\uac{N}+N^a\biggl\{\Gamma_{aIJ} - 2\sigma
	C_{a[I}n_{J]} \notag\\
	&& -2\sigma\ovg{\Gamma}_{a[I|K}n_{|J]}n^K+\sigma\uac{\uac{h}}_{ab}\tilde{\Pi}^b{}_{[I}\tilde{\mathcal{G}}_{J]K}n^K\notag\\
                &&+\frac{1}{4\uac{N}}\uac{\uac{h}}_{ab}N^b\Bigl[\sigma(P^{-1})_{IJKL}\tilde{\mathcal{G}}^{KL}+2n_{[I}
	\tilde{\mathcal{G}}_{J]K}n^K\Bigr]\biggr\}\notag\\
                &&-\frac{1}{4}\uac{N}\left[\tilde{\mathcal{G}}_{IJ}-(P^{-1})_{IJKL}\tilde{\mathcal{G}}^{KL}+2\sigma
	n_{[I}\tilde{\mathcal{G}}_{J]K}n^K\right]. \notag \\ \label{LIJ}
        \end{eqnarray}
	Since $\lambda_{IJ}$ (or $\omega_{tIJ}$), $N^a$ and $\uac{N}$ appear linearly in the action~\eqref{action3}, they play the role of Lagrange multipliers and impose $\tilde{\mathcal{G}}^{IJ}$, $\tilde{\mathcal{D}}_a$ and $\tilde{\tilde{\mathcal{H}}}$ as constraints, respectively. These constraints, known correspondingly as the Gauss, diffeomorphism and scalar constraints, are the same as the ones found in Ref.~\cite{Montesinos1801} by solving the second-class constraints of general relativity in a manifestly $SO(3,1)$ [or $SO(4)$] covariant fashion. Therefore, by following a different path along which the introduction of second-class constraints in the theory is utterly avoided, we have arrived at the same Hamiltonian formulation of general relativity. It is worth mentioning that in Ref.~\cite{Montesinos1801} there is a sign ambiguity $\epsilon$ in the solution of the second-class constraints (since they are quadratic in the canonical momenta) that later propagates along the canonical analysis; such an ambiguity was completely avoided in the present work because no second-class constraints were introduced here. In particular, the projector $W_a{}^b{}_{IJK}$ acting on $\ovg{\omega}_b{}^{JK}$ in Eq. (8) of Ref.~\cite{Montesinos1801} carries and $\epsilon$, while in~\eqref{defC} and~\eqref{W} of the current paper there is no such an $\epsilon$.
	
		To sum it up, from the initial 16 free variables contained in the orthonormal frame $e_{\mu}{}^I$, four of them, associated to the time component $e_{t}{}^I$, play the role of Lagrange multipliers that impose the diffeomorphism and scalar constraints, $\tilde{\mathcal{D}}_a\approx0$ and $\tilde{\tilde{\mathcal{H}}}\approx0$ respectively, whereas the remaining 12 components $e_a{}^I$ of the frame are absorbed into the canonical variable $\tilde{\Pi}^{aI}$, which is related to them by~\eqref{Pi} or~\eqref{eaI}. On the other hand, from the initial 24 variables in the connection $\omega_{\mu IJ}$, the six components $\omega_{t IJ}$ are involved in the Lagrange multipliers $\lambda_{IJ}$ that impose the Gauss constraint $\tilde{\mathcal{G}}^{IJ}\approx0$, the six variables $\uac{\lambda}_{ab}$ are auxiliary fields fixed by their own equation of motion and given by~\eqref{fixlambd}, and the remaining 12 variables $C_{aI}$ constitute the configuration variables that, together with $\tilde{\Pi}^{aI}$, make up the canonical variables of the theory; according to~\eqref{action3}, they are normalized such that the fundamental Poisson bracket reads $\{C_{aI}(t,x),\tilde{\Pi}^{bJ}(t,y)\}=(1/2\kappa)\delta_a^b\delta_I^J\delta^3(x,y)$, where $\delta^3(x,y)$ is the three-dimensional Dirac delta. 
	
	Also, notice that in the case of a vanishing cosmological constant ($\Lambda=0$), the formulation described by the action \eqref{action3} is invariant, up to a global factor, under a constant rescaling of the momenta variables $\tilde{\Pi}^{aI} \rightarrow \Omega \tilde{\Pi}^{aI}$, with $\Omega$ being a nonvanishing real number, since both the internal vector $n_{I}$ and the connection $\Gamma_{aIJ}$ are left invariant by this change [see~\eqref{vectnpi} and~\eqref{GaIJ}, respectively]. Thus, rescaling $\tilde{\Pi}^{aI}$ together with a redefinition of the Lagrange multipliers $\lambda_{IJ} \rightarrow \Omega^{-1} \lambda_{IJ}$, $N^{a} \rightarrow \Omega^{-1} N^{a}$, and $\uac{N} \rightarrow \Omega^{-2} \uac{N}$, leaves the action \eqref{action3} almost unaltered, because the constraints \eqref{Gauss2}, \eqref{diff}, and \eqref{scalar3} remain the same; however, the theory now obeys the fundamental Poisson bracket $\{C_{aI}(t,x),\tilde{\Pi}^{bJ}(t,y)\}=(1/2\kappa\Omega)\delta_a^b\delta_I^J\delta^3(x,y)$. This property has already been exploited within the time gauge framework ~\cite{ThieBook}. Here, we just showed that it is a distinctive feature of the Hamiltonian formulation of general relativity without a cosmological constant, regardless of any gauge fixation.
	
	Furthermore, we also point out that it is not necessary to split $e_{t}{}^I$ into lapse and shift components, which provides a way of unifying the vector and scalar constraints into one $SO(3,1)$ [or $SO(4)$] covariant constraint that, up to terms proportional to the Gauss constraint, takes the form,
	\begin{equation}
		\tilde{\mathcal{H}}_I:=h^{-1/4}\bigl(2\tilde{\Pi}^a{}_I\tilde{\mathcal{D}}_a+\sigma n_I\tilde{\tilde{\mathcal{H}}}\bigr).\label{Unified}
	\end{equation}
	Hence, $\tilde{\mathcal{H}}_I$ and $\tilde{\mathcal{G}}^{IJ}$ constitute the only constraints of the theory, and whereas the latter generates local $SO(3,1)$ [or $SO(4)$] transformations, the former is related to spacetime diffeomorphisms.
	
	\section{Time gauge}\label{TG}
	
	The time gauge fixes the freedom to perform boost transformations and leaves a remnant $SO(3)$ gauge symmetry. The time gauge is imposed by hand through the constraint $\tilde{\Pi}^{a0}\approx0$, which weakly commutes--in Dirac's sense~\cite{dirac1964lectures}--with all the constraints except with $\tilde{\mathcal{G}}^{i0}\approx0$ (boost generator), for which the Poisson bracket gives
	\begin{equation}
		\{\tilde{\Pi}^{a0}(t,x),\tilde{\mathcal{G}}^{i0}(t,y)\}=-\frac{\sigma}{2\kappa}\tilde{\Pi}^{ai}\delta^3 (x,y). 
	\end{equation}
	This renders the pair $(\tilde{\Pi}^{a0},\tilde{\mathcal{G}}^{i0})$ second class because $\tilde{\Pi}^{ai}$ is an invertible $3\times3$ matrix that is associated with the densitized triad through~\eqref{Pi}. We make the second-class constraints strongly equal to zero. From~\eqref{Gauss2}, the solution of  $\tilde{\mathcal{G}}^{i0}=0$ is
	\begin{equation}
		C_{a0}=-\sigma n^0\uac{\Pi}_{ai}\Gamma_b{}^i{}_j\tilde{\Pi}^{bj},\label{Ca0}
	\end{equation}
	where $\uac{\Pi}_{ai}$ is the inverse of $\tilde{\Pi}^{ai}$. Likewise, the time gauge implies $n_0=\text{sgn}(\det(\tilde{\Pi}^{ai}))$ and $n_i=0$ from~\eqref{vectnpi}, and $\Gamma_{a0i}=0$ from~\eqref{GaIJ}. Moreover, the $SO(3)$ indices are raised and lowered with the Euclidean metric $\delta_{ij}$.
	
	Let us define the $SO(3)$ connection $\Gamma_{ai}:=-(1/2)\epsilon_{ijk}\Gamma_a{}^{jk}$; Eq.~\eqref{covdev} then implies that $\Gamma_{ai}$ is the connection compatible with $\tilde{\Pi}^{ai}$: $\nabla_a \tilde{\Pi}^{bi}=\partial_a \tilde{\Pi}^{bi}+\Gamma^b{}_{ac}\tilde{\Pi}^{ci}-\Gamma^c{}_{ac}\tilde{\Pi}^{bi}+\epsilon^{ijk}\Gamma_{aj}\tilde{\Pi}^b{}_{k}=0$. Its curvature is given by $R_{abi}:=-(1/2)\epsilon_{ijk}R_{ab}{}^{jk}=\partial_a \Gamma_{bi}-\partial_b \Gamma_{ai}+\epsilon_{ijk}\Gamma_{a}{}^{j} \Gamma_{b}{}^k$, and it describes the intrinsic geometry of $\Sigma$. According to~\eqref{GaIJ}, $\Gamma_{ai}$ is given explicitly by
	\begin{equation}
	\Gamma_{ai} = -\epsilon_{i j k}\left(\partial_{[b}\underaccent{\tilde}{\Pi}_{a]}{}^{j} + \underaccent{\tilde}{\Pi}_{a}{}^{[l|}\tilde{\Pi}^{c|j]}\partial_{b}
	\underaccent{\tilde}{\Pi}_{cl}\right)\tilde{\Pi}^{bk}.
	\end{equation}

	Therefore, in the time gauge, the action~\eqref{action3} reduces to
	\begin{eqnarray}
		S=&&\kappa\int_{\mathbb{R}\times\Sigma} dtd^3x \left (2\tilde{\Pi}^{ai}\dot{C}_{ai}-2\lambda_i\tilde{\mathcal{G}}^i-2N^a\tilde{\mathcal{D}}_a-\uac{N}\tilde{\tilde{\mathcal{H}}} \right ), \nonumber\\
		&& \label{actionTG}
	\end{eqnarray}
	where $\lambda_i:=-(1/2)\epsilon_{ijk}\lambda^{jk}$. The $SO(3)$ Gauss constraint $\tilde{\mathcal{G}}_i:=-(1/2)\epsilon_{ijk}\tilde{\mathcal{G}}^{jk}$ and the diffeomorphism and scalar constraints take the form,
		\begin{subequations}
		\begin{eqnarray}
		\tilde{\mathcal{G}}_i &=& -\frac{n^0}{\gamma}\left[\partial_a\tilde{\Pi}^a{}_i+\epsilon_{ijk}(-n^0\gamma C_a{}^j)\tilde{\Pi}^{ak}\right],\label{Gaussrot}\\
		\tilde{\mathcal{D}}_a &=& 2\tilde{\Pi}^{bi}\partial_{[a}C_{b]i}-C_{ai}\partial_b\tilde{\Pi}^{bi},\label{diffC}\\
		\tilde{\tilde{\mathcal{H}}} &=& \sigma\epsilon_{ijk}\tilde{\Pi}^{ai}\tilde{\Pi}^{bj} R_{ab}{}^{k}+ 2 \sigma n^0 \Lambda  \det(\tilde{\Pi}^{ai})\notag\\
		&& + 2\tilde{\Pi}^{a[i|}\tilde{\Pi}^{b|j]}\!\left[C_{ai}\!+\!\frac{n^0}{\gamma}\Gamma_{ai}\right]\!\left[C_{bj}\!+\!\frac{n^0}{\gamma}\Gamma_{bj}\right].\label{scalarC}
		\end{eqnarray}
		\end{subequations}
	From~\eqref{Gaussrot} we infer that the object $A_{ai}:=-n^0\gamma C_{ai}$ is an $SO(3)$ connection and we can define its field strength by $F_{abi}:=\partial_{a} A_{bi}-\partial_{b} A_{ai}+\epsilon_{ijk}A_{a}{}^{j} A_{b}{}^k$. In terms of the connection $A_{ai}$, the action~\eqref{actionTG} reads
	\begin{eqnarray}
		S=&&\kappa\int_{\mathbb{R}\times\Sigma}dtd^3x \Biggl( - \frac{2}{\gamma} n^0 \tilde{\Pi}^{ai}\dot{A}_{ai}-2\lambda_i\tilde{\mathcal{G}}^i-2N^a\tilde{\mathcal{D}}_a\notag\\
		&&-\uac{N}\tilde{\tilde{\mathcal{H}}} \Biggr),\label{actionTG2}
	\end{eqnarray}	
	with	
	\begin{subequations}
		\begin{eqnarray}
		    \tilde{\mathcal{G}}_i &=& -\frac{n^0}{\gamma}\left[\partial_a\tilde{\Pi}^a{}_i+\epsilon_{ijk} A_a{}^j \tilde{\Pi}^{ak}\right],\label{gauss0}\\
			\tilde{\mathcal{D}}_a &=& -\frac{n^0}{\gamma}  \left( 2\tilde{\Pi}^{bi}\partial_{[a}A_{b]i}-A_{ai}\partial_b\tilde{\Pi}^{bi} \right),\label{diff0}\\
			\tilde{\tilde{\mathcal{H}}} &=&  \dfrac{1}{ \gamma^{2}}\epsilon_{ijk}\tilde{\Pi}^{ai}\tilde{\Pi}^{bj} \left[ F_{ab}{}^{k} + \left( \sigma \gamma^{2}-1 \right) R_{ab}{}^{k} \right] \notag \\
			& & + 2 \sigma n^0 \Lambda  \det(\tilde{\Pi}^{ai}) - 2 \frac{n^0}{\gamma}  \tilde{\Pi}^a{}_i \nabla_a \tilde{\mathcal{G}}^i,\label{scalarbarb0}
		\end{eqnarray}
	\end{subequations}
	where we have used the identity,
	\begin{eqnarray}
	\epsilon_{ijk}&&(A_a{}^j-\Gamma_a{}^j)(A_b{}^k-\Gamma_b{}^k) \notag \\
	&&=F_{abi}-R_{abi}-2\nabla_{[a}(A_{b]i}-\Gamma_{b]i})\label{relcurv}
	\end{eqnarray}
	to rewrite the last term of~\eqref{scalarC}. Integrating by parts the last term in $\tilde{\tilde{\mathcal{H}}}$ and redefining the Lagrange multiplier in front of the Gauss constraint as $\mu_i:=\lambda_i+(n^0/\gamma)  \tilde{\Pi}^a{}_i \nabla_a\uac{N}$, we get
	\begin{eqnarray}
	S=&&\kappa\!\int_{\mathbb{R}\times\Sigma}\!dtd^3x \Biggl( \!-\frac{2}{\gamma}n^0\tilde{\Pi}^{ai}\dot{A}_{ai}\!-\!2\mu_i\tilde{\mathcal{G}}^i-2N^a\tilde{\mathcal{D}}_a-\uac{N}\tilde{\tilde{\mathcal{C}}} \Biggr), \nonumber\\
	&& \label{actionTG3}
	\end{eqnarray}
	with
	\begin{eqnarray}
	\tilde{\tilde{\mathcal{C}}} &:=& \dfrac{1}{ \gamma^{2}}\epsilon_{ijk}\tilde{\Pi}^{ai}\tilde{\Pi}^{bj} \left[ F_{ab}{}^{k} + \left( \sigma \gamma^{2}-1 \right) R_{ab}{}^{k} \right] \notag \\
			&& + 2 \sigma n^0 \Lambda  \det(\tilde{\Pi}^{ai}).\label{scalarbarb00}
	\end{eqnarray}
	Thereby, we have straightforwardly arrived at the Ashtekar-Barbero formulation for general relativity with cosmological constant~\cite{Barbero9505, Holst} (see also Ref.~\cite{ThieBook}). From~\eqref{actionTG3}, we can read off the Poisson bracket $\{A_{ai}(t,x),\tilde{\Pi}^{bj}(t,y)\}=(-n^0\gamma/2\kappa)\delta_a^b\delta_i^j\delta^3(x,y)$. Notice that, in order to coincide with the results of Ref.~\cite{Holst}, we must take $n^0=-1$ in the Lorentzian case; this amounts to taking $\det(\Pi^{ai})>0$, as in the analysis carried out by Holst.
	
	Alternatively, by using again~\eqref{relcurv} we can get rid of the term involving $R_{ab}{}^i$ in the scalar constraint \eqref{scalarbarb00}, to get
	\begin{eqnarray}
	\tilde{\tilde{\mathcal{C}}}=\tilde{\tilde{\mathcal{S}}} - 2n^0 \left(\sigma\gamma-\frac{1}{\gamma}\right) \tilde{\Pi}^{a}{}_i\nabla_a \tilde{\mathcal{G}}^i, \label{xxx}
	\end{eqnarray}
	with
	\begin{eqnarray}
	\tilde{\tilde{\mathcal{S}}} &:=&  \sigma\epsilon_{ijk}\tilde{\Pi}^{ai}\tilde{\Pi}^{bj} \biggl[ F_{ab}{}^{k} - \left( 1-\frac{\sigma}{\gamma^{2}} \right)\epsilon^{klm} (A_{al}-\Gamma_{al}) \notag \\
	&& \times (A_{bm}-\Gamma_{bm})\biggr]  + 2 \sigma n^0 \Lambda  \det(\tilde{\Pi}^{ai}).
	\label{scalarbarb2.1}
	\end{eqnarray}
Substituting~\eqref{xxx} into the action~\eqref{actionTG3}  and integrating by parts the last term in~\eqref{xxx}, we get
\begin{eqnarray}
	S=&&\kappa\int_{\mathbb{R}\times\Sigma}dtd^3x \Big(- \frac{2}{\gamma} n^0 \tilde{\Pi}^{ai}\dot{A}_{ai}-2\rho_i\tilde{\mathcal{G}}^i-2N^a\tilde{\mathcal{D}}_a \notag \\
	&& - \uac{N} \tilde{\tilde{\mathcal{S}}}  \Big),  \label{actionTG4}
	\end{eqnarray}
	where $\rho_i:=\mu_i +  n^0 (\sigma \gamma - \gamma^{-1}) \tilde{\Pi}^a{}_i \nabla_a\uac{N}$. This alternative form of the scalar constraint agrees with the one reported in Ref.~\cite{ThieBook}. 
	
	As usual, instead of the diffeomorphism constraint $\tilde{\mathcal{D}}_a$, we can use the vector constraint,
	\begin{eqnarray}
	\tilde{\mathcal{C}}_a :=\tilde{\mathcal{D}}_a + A_{ai}  \tilde{\mathcal{G}}^i =  -\frac{n^0}{\gamma}\tilde{\Pi}^{bi}F_{abi}, \label{vector}
	\end{eqnarray}
in the previous actions~\eqref{actionTG2},~\eqref{actionTG3}, and~\eqref{actionTG4} by redefining the corresponding Lagrange multiplier enforcing the 
Gauss constraint.
		
	\section{Other manifestly $\boldsymbol{SO(3,1)}$ [or $\boldsymbol{SO(4)}$] covariant canonical variables}\label{other}
	
	As shown in our previous work~\cite{Montesinos1801}, the manifestly $SO(3,1)$ [or $SO(4)$] covariant formulation of the Holst action contained in~\eqref{action3} of this paper can, alternatively, be expressed in terms of other manifestly $SO(3,1)$ [or $SO(4)$] covariant canonical variables. The underlying canonical transformations are such that they leave the canonical momenta $\tilde{\Pi}^{aI}$ unchanged, whereas the configuration variables are promoted to new ones. They can be encompassed by the map $(C_ {aI},\tilde{\Pi}^{aI})\mapsto(X_ {aI},\tilde{\Pi}^{aI})$, where $X_ {aI}$ is given by
	\begin{eqnarray}\label{cantransf}
		X_{aI}=C_{aI}-W_a{}^b{}_{IJK}\left(\alpha\Gamma_b{}^{JK}+\frac{\beta}{\gamma}\ast\Gamma_b{}^{JK}\right),
	\end{eqnarray}
	where $\alpha$ and $\beta$ are continuous real parameters. The transformation is indeed canonical, since the canonical symplectic structure in~\eqref{action3} changes by a boundary term,
	\begin{eqnarray}
		2\tilde{\Pi}^{aI}\dot{C}_{aI} &=& 2\tilde{\Pi}^{aI}\dot{X}_{aI} +\partial_a \Big(-2\alpha n_I\dot{\tilde{\Pi}}{}^{aI} \notag\\
		&& +\frac{\sigma\beta}{\gamma}\sqrt{h}\tilde{\eta}^{abc}\uac{\uac{h}}_{bd}\uac{\uac{h}}_{cf}\dot{\tilde{\Pi}}^{dI}\tilde{\Pi}^f{}_I \Big).
	\end{eqnarray}
	In the new variables, the action~\eqref{action3} acquires the form,
	\begin{eqnarray}
	S=&&\kappa\int_{\mathbb{R}\times\Sigma}dtd^3x \Big( 2\tilde{\Pi}^{aI}\dot{X}_{aI}-\lambda_{IJ}\tilde{\mathcal{G}}^{IJ}-2N^a\tilde{\mathcal{D}}_a \notag\\
	&& -\uac{N}\tilde{\tilde{\mathcal{H}}} \Big), \label{action5}
	\end{eqnarray}
	where the Gauss, diffeomorphism and scalar constraints are given by
	\begin{widetext}
	\begin{subequations}
		\begin{eqnarray}
		\tilde{\mathcal{G}}^{IJ} &=& 2\tilde{\Pi}^{a[I}X_a{}^{J]}+4\left[(1-\alpha)\delta_{[K}^I\delta_{L]}^J +\frac{(1-\beta)}{2\gamma}\epsilon^{IJ}{}_{KL}\right]\tilde{\Pi}^{a[K}n^{M]}\Gamma_a{}^L{}_M,\label{Gausscan}\\
		\tilde{\mathcal{D}}_a &=& 2\tilde{\Pi}^{bI}\partial_{[a}X_{b]I}-X_{aI}\partial_b\tilde{\Pi}^{bI},\label{diffcan}\\
		\tilde{\tilde{\mathcal{H}}} &=&-\sigma\tilde{\Pi}^{aI}\tilde{\Pi}^{bJ}R_{abIJ} +2\tilde{\Pi}^{a[I|}\tilde{\Pi}^{b|J]}\Biggl\{X_{aI}X_{bJ}+\left(\frac{1-\beta}{\gamma}\right)^2q^{KL}\Gamma_{aIK}\Gamma_{bJL}+2X_{aI}\biggl[(1-\alpha)\Gamma_{bJK}+\frac{(1-\beta)}{\gamma}\ast\Gamma_{bJK}\biggr] n^K \nonumber\\
		&& + (1-\alpha)\left[(1-\alpha)\Gamma_{aIK}+\frac{2}{\gamma}(1-\beta)\ast\Gamma_{aIK}\right]\Gamma_{bJL}n^Kn^L\Biggr\}+2\sigma\Lambda \sqrt{h},\label{scalarcan}
		\end{eqnarray}
	\end{subequations}
	\end{widetext}
respectively. Notice that the diffeomorphism constraint takes exactly the same form as the original one and that it is independent of both $\alpha$ and $\beta$; this means that $X_{aI}$ transforms as a 1-form under spatial diffeomorphisms for any choice of $\alpha$ and $\beta$. On the other hand, the Gauss and scalar constraints strongly depend on the values of these parameters. However, if $\beta=1$ the Immirzi parameter $\gamma$ drops out from~\eqref{Gausscan} and~\eqref{scalarcan} regardless of the value of $\alpha$. We conclude by analyzing some particular nontrivial cases of the canonical transformation~\eqref{cantransf} ($\alpha=0=\beta$ is the identity transformation),
\begin{itemize}
	\item[(i)] For $\alpha=1=\beta$ the configuration variable $X_{aI}$ becomes the configuration variable $Q_{aI}$ introduced in Ref.~\cite{Montesinos1801}. The Gauss and scalar constraints simplify considerably in terms of the phase-space variables $(Q_{aI}, \tilde{\Pi}^{aI})$. Notice that the Immirzi parameter does not appear in the constraints, which take exactly the same form as those arising in the manifestly $SO(3,1)$ [or $SO(4)$] covariant canonical analysis of the Palatini action~\cite{Peldan9400}. 
	
	\item[(ii)] For $\alpha=1$ and $\beta=0$ the configuration variable $X_{aI}$ becomes the configuration variable $K_{aI}$ introduced in Ref.~\cite{Montesinos1801}. The Gauss and scalar constraints simplify a bit in terms of the phase-space variables $(K_{aI}, \tilde{\Pi}^{aI})$, but the Immirzi parameter is still present.
	
	\item[(iii)] For $\alpha=0$ and $\beta=1$ the configuration variable $X_{aI}$ will be denoted as ${\cal Q}_{aI}$. This case had not been previously reported in literature and leads to new phase-space variables $({\cal Q}_{aI}, \tilde{\Pi}^{aI})$. In terms of them, the constraints \eqref{Gausscan}-\eqref{scalarcan} take the form,
	\begin{subequations}
		\begin{eqnarray}
		\tilde{\mathcal{G}}^{IJ} &=& 2\tilde{\Pi}^{a[I} {\cal Q}_a{}^{J]}+ 2 \tilde{\Pi}^{a[I} \Gamma_a{}^{J]}{}_M n^M \notag\\
		&& - 2 \tilde{\Pi}^{aM} n^{[I} \Gamma_a{}^{J]}{}_M,\label{GausscanS}\\
		\tilde{\mathcal{D}}_a &=& 2\tilde{\Pi}^{bI}\partial_{[a} {\cal Q}_{b]I} - {\cal Q}_{aI}\partial_b\tilde{\Pi}^{bI},\label{diffcanS}\\
		\tilde{\tilde{\mathcal{H}}} &=& -\sigma\tilde{\Pi}^{aI}\tilde{\Pi}^{bJ}R_{abIJ} +2\tilde{\Pi}^{a[I|}\tilde{\Pi}^{b|J]}\biggl( {\cal Q}_{aI} {\cal Q}_{bJ}\nonumber\\
		&&+ 2 {\cal Q}_{aI}\Gamma_{bJK} n^K +\Gamma_{aIK}\Gamma_{bJL}n^Kn^L\biggr) \notag\\
		&& +2\sigma\Lambda \sqrt{h}. \label{scalarcanS}
		\end{eqnarray}
	\end{subequations}
\end{itemize}
	
	Notice that these constraints are independent of the Immirzi parameter too. Hence, the configuration variables ${\cal Q}_{aI}$--together with the configuration variables $Q_{aI}$--are naturally associated with the Palatini action~\cite{Montesinosprog1}. The canonical transformations (i) and (iii) can be regarded as $SO(3,1)$ [or $SO(4)$] versions of the inverse of Barbero’s canonical transformation.

	\paragraph*{Time gauge.}In the time gauge, the canonical transformation \eqref{cantransf} becomes, after using~\eqref{Ca0},
	\begin{subequations}
		\begin{eqnarray}
		&&X_ {a0}=\sigma n^0(\alpha-1)\tilde{\Pi}^{bi}\partial_b \uac{\Pi}_{ai},\label{Xa0}\\
		&&X_{ai}=C_{ai}+\frac{n^0\beta}{\gamma}\Gamma_{ai}=-\frac{n^0}{\gamma}(A_{ai}-\beta\Gamma_{ai}),\label{barberobeta}
		\end{eqnarray}
	\end{subequations}
	the latter being independent of $\alpha$ (this parameter only appears in the expression found for $X_{a0}$, whose value is not relevant in the time gauge). Moreover, from~\eqref{Gausscan} the $SO(3)$ Gauss constraint, $\tilde{\mathcal{G}}_i =-(1/2)\epsilon_{ijk}\tilde{\mathcal{G}}^{jk}$, becomes
	\begin{eqnarray}\label{GaussX}
		\tilde{\mathcal{G}}_i=\epsilon_{ijk}X_a{}^j\tilde{\Pi}^{ak}-\frac{n^0}{\gamma}(1-\beta)\partial_a\tilde{\Pi}^a{}_i,
	\end{eqnarray}
	whereas the constraints \eqref{diffcan} and \eqref{scalarcan} read
	\begin{subequations}
	\begin{eqnarray}
		\tilde{\mathcal{D}}_a &=& 2\tilde{\Pi}^{bi}\partial_{[a}X_{b]i}-X_{ai}\partial_b\tilde{\Pi}^{bi},\label{diffX}\\
		\tilde{\tilde{\mathcal{H}}} &=& \sigma\epsilon_{ijk}\tilde{\Pi}^{ai}\tilde{\Pi}^{bj} R_{ab}{}^{k}+ 2 \sigma n^0 \Lambda  \det(\tilde{\Pi}^{ai}) + 
		2\tilde{\Pi}^{a[i|}\tilde{\Pi}^{b|j]} \notag\\
		&& \times  \left[ X_{ai}\!+\!\frac{n^0}{\gamma}(1-\beta)\Gamma_{ai}\right]\!\left[X_{bj}\!+\!\frac{n^0}{\gamma}(1-\beta)\Gamma_{bj} \right],\notag\\
		\label{scalarX}
	\end{eqnarray}
	\end{subequations}
	respectively. Therefore, the action \eqref{action5} takes the form,
	\begin{eqnarray}
	&&S=\kappa\int_{\mathbb{R}\times\Sigma}dtd^3x \Big( 2\tilde{\Pi}^{ai}\dot{X}_{ai}-2\lambda_i\tilde{\mathcal{G}}^i-2N^a\tilde{\mathcal{D}}_a-\uac{N}\tilde{\tilde{\mathcal{H}}}  \Big). \nonumber\\
	&& \label{action6}
	\end{eqnarray}
	Notice that the action is independent of $\alpha$, whereas the value of $\beta$ determines the canonical theory under consideration, according to whether $\beta=1$ or $\beta\neq1$. Let us analyze this in detail,
	
	\begin{itemize}
		\item[(a)] For $\beta=1$, the constraints \eqref{GaussX}, \eqref{diffX} and \eqref{scalarX} become 
		\begin{subequations}
			\begin{eqnarray}
			\tilde{\mathcal{G}}_i &=& \epsilon_{ijk}X_a{}^j\tilde{\Pi}^{ak},\label{gaussADM}\\
			\tilde{\mathcal{D}}_a &=& 2\tilde{\Pi}^{bi}\nabla_{[a}X_{b]i}-\Gamma_{ai}\tilde{\mathcal{G}}^i\approx 2\tilde{\Pi}^{bi}\nabla_{[a}X_{b]i},\\
			\tilde{\tilde{\mathcal{H}}} &=& \sigma\epsilon_{ijk}\tilde{\Pi}^{ai}\tilde{\Pi}^{bj} R_{ab}{}^{k}+2\tilde{\Pi}^{a[i|}\tilde{\Pi}^{b|j]}X_{ai}X_{bj}\notag\\
			&&+ 2 \sigma n^0 \Lambda  \det(\tilde{\Pi}^{ai}),\label{scalarADM}
			\end{eqnarray}
		\end{subequations}
		which can be recognized as the constraints of the $SO(3)$ ADM formalism~\cite{ashtekar1991lectures}. Note also that~\eqref{barberobeta} is the converse of Barbero's transformation since we go from the connection $A_{ai}$ to the vector $X_{ai}$. This means that, for $\beta=1$ and regardless of the value of $\alpha$, we obtain the $SO(3)$ ADM formulation of general relativity, where $-n^0X_{ai}$ is an object closely related to the extrinsic curvature. As already mentioned, the configuration variables ${\cal Q}_{aI}$ ($\alpha=0$) and $Q_{aI}$ ($\alpha=1$) belong to the case $\beta=1$. Therefore, from \eqref{Xa0} and \eqref{barberobeta}, they are set to ${\cal Q}_{a0}=-\sigma n^0\tilde{\Pi}^{bi}\partial_b \uac{\Pi}_{ai}$, $Q_{a0}=0$ and ${\cal Q}_{ai}=-(n^0/\gamma)(A_{ai}-\Gamma_{ai})=Q_{ai}$ in the time gauge.
		
		\item[(b)] For $\beta\neq 1$, note that~\eqref{GaussX}, \eqref{diffX} and \eqref{scalarX} resemble~\eqref{Gaussrot}-\eqref{scalarC} with $\gamma/(1-\beta)$ taking the place of $\gamma$. In fact, we can express the Gauss constraint~\eqref{GaussX} as
		\begin{eqnarray}
		\tilde{\mathcal{G}}_i=-\frac{n^0}{\gamma}(1-\beta)\left(\partial_a\tilde{\Pi}^a{}_i+\epsilon_{ijk}\mathcal{A}_a{}^j\tilde{\Pi}^{ak}\right),\label{GaussX0}
		\end{eqnarray}
		where we have identified the $SO(3)$ connection $\mathcal{A}_{ai}:=-n^0\gamma X_{ai}/(1-\beta)$. Let us denote its field strength by $\mathcal{F}_{abi}:=\partial_{a} \mathcal{A}_{bi}-\partial_{b} \mathcal{A}_{ai}+\epsilon_{ijk}\mathcal{A}_{a}{}^{j} \mathcal{A}_{b}{}^k$. Then, using the identity,
		\begin{eqnarray}
		\epsilon_{ijk}&&(\mathcal{A}_a{}^j-\Gamma_a{}^j)(\mathcal{A}_b{}^k-\Gamma_b{}^k)  \notag \\ 
		&& = \mathcal{F}_{abi}-R_{abi}-2\nabla_{[a}(\mathcal{A}_{b]i}-\Gamma_{b]i}), \label{idxxx}
		\end{eqnarray}
		we rewrite the last term of~\eqref{scalarX}, so the action \eqref{action6} becomes
		\begin{eqnarray}
		S=&&\kappa\int_{\mathbb{R}\times\Sigma}dtd^3x \Big[ - \frac{2}{\gamma} n^0(1-\beta) \tilde{\Pi}^{ai}\dot{\mathcal{A}}_{ai} \notag\\
		&&-2\lambda_i\tilde{\mathcal{G}}^i -2N^a\tilde{\mathcal{D}}_a-\uac{N}\tilde{\tilde{\mathcal{H}}} \Big], \label{actionTX2}
		\end{eqnarray}	
		with	
		\begin{subequations}
			\begin{eqnarray}
			\tilde{\mathcal{D}}_a &=& -\frac{n^0}{\gamma}(1-\beta)  \left( 2\tilde{\Pi}^{bi}\partial_{[a}\mathcal{A}_{b]i}-\mathcal{A}_{ai}\partial_b\tilde{\Pi}^{bi} \right),\label{diffX0}\\
			\tilde{\tilde{\mathcal{H}}} &=& \dfrac{(1-\beta)^2}{ \gamma^{2}}\epsilon_{ijk}\tilde{\Pi}^{ai}\tilde{\Pi}^{bj} \notag \\
			&& \times \left\{ \mathcal{F}_{ab}{}^{k} + \left[ \frac{\sigma\gamma^{2}}{(1-\beta)^2}-1 \right] R_{ab}{}^{k} \right\} \notag \\
			&& + 2 \sigma n^0 \Lambda  \det(\tilde{\Pi}^{ai}) - 2 \frac{n^0}{\gamma} (1-\beta)  \tilde{\Pi}^a{}_i \nabla_a \tilde{\mathcal{G}}^i. \notag \\
			\label{scalarbarbX0}
			\end{eqnarray}
		\end{subequations}
		Integrating by parts the last term of~\eqref{scalarbarbX0}, the action~\eqref{actionTX2} finally acquires the form,
		\begin{eqnarray}
		S=&&\kappa\!\int_{\mathbb{R}\times\Sigma}\!dtd^3x \Big[ -\frac{2}{\gamma}n^0(1-\beta)\tilde{\Pi}^{ai}\dot{\mathcal{A}}_{ai}\!\notag\\
		&&-\!2\nu_i\tilde{\mathcal{G}}^i-2N^a\tilde{\mathcal{D}}_a -\uac{N}\tilde{\tilde{\mathcal{C}}} \Big], \label{action7}
		\end{eqnarray}
		where $\nu_i:=\lambda_i + [n^0(1-\beta)/\gamma] \tilde{\Pi}^a{}_i \nabla_a\uac{N}$ and
		\begin{eqnarray}
		\tilde{\tilde{\mathcal{C}}} &=& \dfrac{(1-\beta)^2}{ \gamma^{2}}\epsilon_{ijk}\tilde{\Pi}^{ai}\tilde{\Pi}^{bj} \notag \\
		&& \times \left\{ \mathcal{F}_{ab}{}^{k} + \left[ \frac{\sigma\gamma^{2}}{(1-\beta)^2}-1 \right] R_{ab}{}^{k} \right\} \notag \\
		&& +2 \sigma n^0 \Lambda  \det(\tilde{\Pi}^{ai}).\label{scalarbarbX02}
		\end{eqnarray}
		Notice that the constraints \eqref{GaussX0}, \eqref{diffX0} and \eqref{scalarbarbX02} take exactly the same form as \eqref{gauss0}, \eqref{diff0} and \eqref{scalarbarb00}, respectively, with $A_{ai}$ and $\gamma$ replaced correspondingly with $\mathcal{A}_{ai}$ and $\gamma/(1-\beta)$. Therefore, for $\beta\neq1$ we obtain the Ashtekar-Barbero formulation with a rescaled Immirzi parameter $\gamma/(1-\beta)$. It is worth realizing that the connections $A_{ai}$ and $\mathcal{A}_{ai}$ are related to each other by $\mathcal{A}_{ai}=(1-\beta)^{-1}(A_{ai}-\Gamma_{ai})+\Gamma_{ai}$. Note that they are the same for $\beta=0$, as expected. As already mentioned, the configuration variable $K_{aI}$ ($\alpha=1$ and $\beta=0$) belongs to the case $\beta \neq 1$. Therefore, from \eqref{Xa0} and \eqref{barberobeta}, it is set to $K_{a0}=0$ and $K_{ai}=-n^0 \gamma^{-1} A_{ai}$ in the time gauge.
	\end{itemize}

		\section{Conclusions}
	
	In this paper, we have carried out from scratch the canonical analysis  of the Holst action without introducing second-class constraints. Our strategy consisted in splitting the 18 degrees of freedom contained in the spatial connection $\omega_{aIJ}$, by means of the orthogonal projectors \eqref{W} and \eqref{U}, into 12 dynamical $SO(3,1)$ [or $SO(4)$] covariant variables $C_{aI}$ plus six additional fields $\uac{\lambda}_{ab}$, something explicitly exhibited in~\eqref{connec_par}. The substitution of this expression into the action then implies a reduction of the presymplectic structure of the theory to a symplectic one, a process in which we realize that the fields $\uac{\lambda}_{ab}$ define the null directions of the former and also play the role of auxiliary fields that can be easily integrated out using their own equation of motion. Doing so, the action becomes~\eqref{action3}, in which the canonical pair $(C_{aI},\tilde{\Pi}^{aI})$ is subject to the first-class constraints \eqref{Gauss2}, \eqref{diff} and \eqref{scalar3}, namely the Gauss, diffeomorphism and scalar constraints, respectively. It is really remarkable how our approach simplifies the canonical analysis of general relativity, not only leading to a manifestly $SO(3,1)$ [or $SO(4)$] covariant parametrization of the phase space of the theory (which agrees with the one found by solving second-class constraints~\cite{Montesinos1801}), but also allowing us to keep track of the role played by each one of the original variables involved in the Holst action. Afterwards, we imposed the time gauge and showed that it immediately leads to the Ashtekar-Barbero variables. In that process, the spatial components $-n^0\gamma C_{ai}$ get identified with the Ashtekar-Barbero connection.
	
	In addition, we discussed the set of canonical transformations given by~\eqref{cantransf}  which depends on two parameters $\alpha$ and $\beta$, and relates different $SO(3,1)$ [or $SO(4)$] covariant parametrizations of the phase space of general relativity; some particular cases of them--$(Q_{aI}, \tilde{\Pi}^{aI})$ and $(K_{aI}, \tilde{\Pi}^{aI})$--have already been reported in the literature~\cite{Montesinos1801}. We also reported the new canonical phase-space variables $({\cal Q}_{aI}, \tilde{\Pi}^{aI})$, for which the constraints are given by \eqref{GausscanS}-\eqref{scalarcanS} and turn out to be independent of the Immirzi parameter. In the time gauge, the canonical theories associated to these variables bifurcate into two kinds depending on the value of $\beta$ ($\alpha$ is not important in the time gauge): for $\beta=1$ we obtain the $SO(3)$ ADM formulation of general relativity, whereas for $\beta\neq 1$ we arrive at the Ashtekar-Barbero formulation with a rescaled Immirzi parameter $\gamma/(1-\beta)$.
	
	Although in this paper we have discussed the canonical analysis of the Holst action with respect to a foliation by spacelike hypersurfaces, our strategy can also be adapted to deal with the case of a timelike foliation as well for Lorentzian signature. In that case, the vector $n^I$ is normalized such that $n_In^I=1$ and we can follow a procedure very similar to that described in Sec.~\ref{analysis}. After integrating out the auxiliary fields, we get the same Lorentz-covariant variables and first-class constraints than those obtained in Ref.~\cite{MontRomEscCel} after solving the second-class constraints. Then, by imposing the space gauge there, the Ashtekar-Barbero formulation with gauge group $SU(1,1)$ as well as the $SO(2,1)$ ADM formulation arise. The detailed analysis of all of this will be reported in another paper. In addition, it would be very interesting to further extend our approach to deal with null foliations and compare the resulting canonical theory with that of Ref.~\cite{Alexandrovnull} for the Palatini action, where tertiary constraints arise.
	
	Our procedure stands out for its simplicity, methodology, and economy. Whereas in Dirac's approach the canonical analysis can be cumbersome due to the amount of variables and constraints involved in the formalism, here we have identified from the very beginning the fundamental variables and gotten rid of the superfluous ones in the process. This indeed simplifies the analysis, since we arrive directly at a manifestly $SO(3,1)$ [or $SO(4)$] covariant description of the phase space of general relativity where the only constraints present in the theory are those associated with the two gauge symmetries underlying general relativity. In addition, the constraints take a simpler form than those obtained in nonmanifestly Lorentz-covariant approaches (compare for instance with Refs.~\cite{Barros0100,MonRomCel_Revit}) and the geometrical meaning of the involved canonical variables is clearly established.
	
	We point out that the current theoretical framework applies to the four-dimensional Palatini action as well, as can be seen in Ref.~\cite{Montesinosprog1} or by taking the limit $\gamma \rightarrow \infty$ in the analysis presented in Sec.~\ref{analysis}; actually, the resulting canonical theory coincides with the one emerging from the Holst action after implementing the canonical transformation discussed in Sec.~\ref{other} for the values of the parameters specified there (see also Ref.~\cite{Montesinos1801}). Moreover, our approach can also be used to perform from scratch the canonical analysis of the $n$-dimensional Palatini action with a cosmological constant~\cite{Montesinosprog1} and to study the coupling of matter fields to the Holst (or Palatini) action in the canonical framework (these results will be reported elsewhere too).
	
	Regarding the unified constraint \eqref{Unified} at the end of Sec.~\ref{analysis}, it would be really interesting to compute the gauge algebra satisfied by this constraint together with the Gauss constraint $\tilde{\mathcal{G}}^{IJ}$ and also to establish the relation between the gauge transformations generated by $\tilde{\mathcal{H}}_I$ and the alternative gauge transformations for the Holst action reported in Ref.~\cite{Montesinos1709}.
	
	We think our results can also be useful for investigating the asymptotic behavior of the gravitational field where Lorentz invariance is relevant as well as for the study of gravitational field configurations using numerical methods, where our approach could provide new insights for positing initial value problems in gravity because it preserves Lorentz invariance. 
	
	The technique of our manuscript was also used to do the Hamiltonian analysis of general relativity with Immirzi parameter, expressed as a $BF$ theory supplemented with constraints on the $B$ field, without introducing second-class constraints in the canonical analysis, which was reported in Ref.~\cite{BFnew}.
	
	Finally, although the canonical variables contained in this paper are no longer connection variables--but they give rise to connection variables in the time gauge, as pointed out above--and thus their implementation in the quantum theory is quite nontrivial, our results should motivate the development of new mathematical techniques to potentially use these variables in applications to quantum gravity.

	\acknowledgments
	
	This work was partially supported by Fondo SEP-Cinvestav and by Consejo Nacional de Ciencia y Tecnolog\'ia (CONACyT), M\'exico, Grant No. A1-S-7701.

	\bibliographystyle{apsrev4-1}
	\bibliography{references}

\end{document}